\documentclass[onecolumn,12pt,a4paper]{article}

\usepackage{ametsoc}

\usepackage{lineno}
\usepackage{amsmath}
\usepackage{graphicx}
\usepackage[left=2.5cm, right=2.5cm, top=1.1in,bottom=1.1in]{geometry}
\usepackage{sectsty}

\begin{document}
%

\begin{center}
\LARGE{Rainfall nowcasting by combining radars, microwave links and rain gauges}
\end{center}


\textsc{Blandine Bianchi}\\
\'Ecole Polytechnique F\'ed\'erale de Lausanne, School of Architecture, Civil and Environmental Engineering, Environmental Remote Sensing Laboratory, Lausanne, Switzerland.\\

\textsc{Peter Jan van Leeuwen} 
Departement of Meteorology, University of Reading, Reading, United Kingdom.\\

\textsc{Robin J. Hogan}
Departement of Meteorology, University of Reading, Reading, United Kingdom.\\

\textsc{Alexis Berne}	
\'Ecole Polytechnique F\'ed\'erale de Lausanne,
School of Architecture, Civil and Environmental Engineering,
Environmental Remote Sensing Laboratory, Lausanne, Switzerland.
\textit{(Corresponding author:alexis.berne@epfl.ch)}


\section{Abstract}
The objective of this work is to provide high-resolution rain rate maps at short lead-time forecasts (nowcasts) necessary to anticipate flooding  and properly manage sewage systems in urban areas by combining radars, rain gauges, and operational microwave links, and taking into account their respective uncertainties.
A variational approach (3D-Var) is used to find the best estimate for the rain rate, and its error covariance, from the different rain sensors. Short-term rain rate forecasts are then produced by assuming Lagrangian persistence. A velocity field is obtained from the operational radar-derived rain fields, and the rain rate field is advected using the Total Variance Diminishing (TVD) scheme. 
The error covariance  associated to the estimated rain rate is also propagated, and we use these two in the 3D-Var at the next observation time step. 
This approach can be seen as a Variational Kalman Filter (VKF), in which the covariance of the prior is not constant but dependent on time.  
The proposed approach has been tested using data from 14 rain gauges, 14 microwave links and the operational radar rain product from MeteoSwiss in the area of Z\"{u}rich (Switzerland).
During the applications the assumption of the Lagrangian persistence appears to be valid up to 20 min (a bit longer for stratiform events). During convective events, the algorithm is less powerful and shorter lead times should be considered (i.e., 15 min). Although such lead times are short, they are still useful to various hydrological and outdoor applications.

\section{Introduction}
 
The weather forecast spanning over the next six hours is generally referred to as nowcast \citep{Wilson_BAMS_1998}. 
The critical importance of nowcast is related to areas where safety, infrastructure protection, flood prevention, airport operations and ground traffic control are concerned. Furthermore, the organization of outdoor activities (e.g., fishing), sport events (e.g., strategy and safety in the racing competitions), agricultural activities are usually tightly linked to the accuracy and reliability of nowcasts.
Because of climate change, some parts of the world will be prone to more severe flash floods due to an increase in the intensity of rainfall \citep{Hapuarachchi_2011}.
Flash floods are characterized by excessive rainfall or by rapid release of water resulting in a very limited opportunity for warnings. Changes in demography and the increased urbanisation will lead to a large part of the population being exposed to flash flooding \citep{Hapuarachchi_2011}. Over half of the fatalities due to flash floods involves people driving through flooded intersections. 
Improved forecast a few tens of minute in advance can contribute to better anticipate and mitigate flash foods, hence reducing their potentially dramatic impact.

Flash floods due to excessive rainfall are usually restricted to urban areas or small-scale catchments close to a water body. The development of such events takes places over short periods of time, so the forecasting community has to provide hydrologists with precipitation forecasts at high spatial and temporal resolutions \citep{Collier_2007, Barillec_AWR_2009}.
The use of new models and data types in flash flood forecasting has attracted increasing attention by researchers over the past decades but a lot of uncertainty in flash flood forecasts is linked to predicting future precipitation \citep{Hapuarachchi_2011}. 
There have been increasing calls to accelerate the worldwide development of nowcasting systems to improve flash flood forecast \citep{Penning_2000, Handmer_2001} and to demonstrate the benefits of nowcasting products to weather-sensitive activities like outdoor sporting events \citep{Keenan_BAMS_2003, Wilson_2010}. 
 
The role of rainfall nowcast is extremely important for urban hydrological applications \citep{Fletcher_AWR_2013}.  The variability of the rain rate can have drastic impacts on the runoff mechanism particularly in an urban context \citep{Berne_JH_2004}. The hydrological models require accurate rainfall estimates with the associated uncertainties, and an ideal rainfall input should have a 1--5 min temporal resolution and a 0.1--1 km spatial resolution. 
 
Currently available nowcast models allow us to forecast small meteorological features such as single rain showers and localized thunderstorms using the latest radar or satellite data in a temporal range of around 6-9 hours. These measurements allow a level of accuracy in the forecast for the following few hours that can not be achieved by Numerical Weather Prediction (NWP) models \citep{Lin_GRL_2005}.
While satellites allow us to extrapolate the movement of big cloudy cells during the next 6--9 hours, their spatial resolution is around 50--100 km \citep{Collier_2002}. Radars help forecast the movement of the rainy cells in the next 1--3 hours with a spatial resolution of the order of 1 km \citep{Collier_2002, Liguori_2011}. Radar data are very detailed and provide information about the intensity, shape, speed and direction of movement of the rain cells on a continuous basis. 

The possibility to use radar measurements to characterize the nowcast rainfall has been extensively investigated in previous studies.
\cite{Zawadzki_JAM_1994} defined the predictability as the ability to forecast precipitation by Lagrangian persistence. Their forecast model is a simple translation of the precipitation field with average storm speed with a time resolution of 10~min and a spatial resolution of 4~km. Their results showed that after a time ranging between 40 and 112~min all forecast skill is lost, depending upon individual case.
\cite{Grecu_JH_2000} investigated the use of three different nowcast schemes: Eulerian, Lagrangian and neural network approach, and computed the precipitation field as a function of forecast resolution. They defined the limit of the predictability when the efficiency becomes negative and the correlation between the forecast and observation decreases below 0.5. Their results show that large scale precipitation features were characterized by larger Lagrangian persistence and that precipitation fields are only predictable at scales that exceeded 20~km after 60~min.
\cite{Germann_MWR_2002} defined the predictability of precipitation fields assuming Eulerian and Lagrangian persistence. They found that the range of predictability increases with increasing scale, but they did not deal with data with resolution below 15~min and 4~km.
Since the predictability varies tremendously based on spatial and temporal scales of observed meteorological phenomena,
\cite{Ruzanski_JAMC_2012} investigated the Eulerian and Lagrangian persistence at a smaller scale to quantify the space-time scale dependency of the short-time lead forecast using radar observations. The results showed that short lead-time forecast have a lifetime of about 15~min and 20~min in Eulerian and Lagrangian persistence respectively using high-resolution data: 0.5~km and 1~min. The Lagrangian persistence represented an improvement of about 30\%--40\% (6~min) over Eulerian persistence.
They also showed that the lifetime values corresponding to Lagrangian persistence increase with decreasing spatial and temporal resolution and an improvement of 10\% can be obtained by updating the velocity field.
This supports the fact that the lifetime of larger scale phenomena are longer than those at smaller scales.

As pointed out by \cite{Zawadzki_JAM_1994} the evolution of precipitation is however more complex than the simple Lagrangian persistence, and it is likely that the quantity to be forecasted is precipitation at ground instead of the radar fields used as ground truth.

To this respect, the use of information closer to the ground will conribute to improve the quality of rain rate forecasts. Recently, operational telecommunication microwave links have been shown to provide relevant rain fall information \citep{Messer_Science_2006, Overeem_PNAS_2013}. \cite{Bianchi_JHM_2013b} have proposed a variational approach to combine rain gauge, radar, and microwave link observations to improve the estimation of rain rate at the ground level.

The objective of the present study is to combine measurements from weather radars, rain gauges and operational telecommunication microwave links in order to provide short lead-time forecasts at high spatial and temporal resolutions required for hydrological applications, in particular in an urban context.
To this aim, we implemented a novel method to estimate the rain rate and the associated uncertainty and propagate them for nowcasting, integrating a variational approach \citep{Bianchi_JHM_2013b} and Kalman filtering \citep{Kalman_JBE_1960}.

The paper is organized as follows: Section~2 gives an overview of the rain sensors that are used in this study. Section~3 presents a novel variational Kalman filter framework that we implemented. The data and study area are described in Section~4.  We present and discuss the results in Section~5. The conclusions and some perspectives are provided in Section~6.

\section{Rainfall instruments}
Rain gauges are generally considered accurate because they directly measure the rain rate, but they have a poor spatial representativeness.   Rain gauge measurement errors can be due to the presence of wind, snow or evaporation \citep[e.g.,][]{Nespor_JAOT_1999, Upton_JH_2003, Sieck_WRR_2007}. Their location must be far from the obstacles and from the sources of heat \citep[e.g.,][]{WMO_2008}. Despite these limitations, rain gauges are still able to provide useful estimates of rainfall intensity, especially at the ground level.

Radars have a high spatial coverage, they give a measure of observable quantities on the whole spatial area and reveal the complete structure of the meteorological phenomenon.
The rain intensity $r$ (mmh$^{-1}$) is estimated from the radar reflectivity $Z$ (mm$^6$m$^{-3}$) with the following relationship:
\begin{linenomath}
\begin{equation}
Z=a r^b,
\label{refl}
\end{equation}
\end{linenomath}
where $a$ and $b$ are subject to uncertainties mainly due to the variability of the raindrop size distribution. The other main sources of uncertainty that can affect radar rain-rate estimates are vertical variability, bright band contamination, ground clutter and attenuation (at C and X band). Operational services have developed and operationally implemented quality control and correction procedures to minimize these uncertainties \citep[e.g.,][]{Germann_QJRMS_2006}.

Microwave links are alternative tools for rain rate estimation \citep{Messer_Science_2006, Leijnse_WRR_2007_c, Overeem_PNAS_2013}.
Using the specific attenuation due to rain affecting the microwave signals, it is possible to get an estimate of the path-averaged rain rate  $r$~(mm h$^{-1}$) through the following power-law equation  \citep{Atlas_JAM_1977, Olsen_IEEE_TAP_1978}:
\begin{linenomath}
\begin{equation}
k=\alpha r^\beta,
\label{att}
\end{equation}
\end{linenomath}
where $k$ (expressed in dB km$^{-1}$) is the specific attenuation affecting the microwave signal and the parameters $\alpha$ and $\beta$ depend on the frequency, polarization, drop size distribution and temperature. Similarly to radar, microwave links provide indirect measurements of rain rate. Other sources of uncertainty are the atmospheric-gas and wet-antenna attenuation, link precision and electronics \citep{ Overeem_WRR_2011, Bianchi_JH_2013a}.

\section{A novel variational Kalman filter framework}

The propagation of the rain field in the near future requires 3D-VAR model, which measures the current rain rate across space, to be upgraded to incorporate the temporal dimension (it should be noted that the problem deals with a 2D spatial domain despite the name 3D-VAR).
In 3D-VAR the cost function, which describes the misfit between the true state and the background as well as the observations, is directly minimized using numerical schemes (e.g., Gauss-Newton).
The standard upgrade to 3D-VAR for forecast purposes is 4D-VAR. 3D-VAR and 4D-VAR usually require the development of an adjoint model that is rigorous but practical implementation raises a number of serious problems because it is computationally much more expensive, even if it allows for assimilation of asynoptic data. 
Operational weather centres have implemented 3D-VAR and 4D-VAR \citep{Rabier_2000, Gauthier_MWR_2005}.
If the model is perfect and linear and the observation operator is also linear, 4D-VAR and Kalman Filter (KF) give the same solution \citep{Bennett_1992}.
The fundamental difference between the Kalman Filter and 4D-VAR lies in how the covariance matrix evolves. 
The KF explicitly calculates the error covariances through an additional matrix that propagates the error information from one update to the next accounting for uncertain model dynamics \citep{Reichle_AWR_2008}.
4D-VAR, instead, does not propagate error covariance information from one assimilation window to the other. It is assumed to be constant, moreover the system is often assumed to be perfect \citep{Rabier_2005}.  
The Ensemble Kalman Filter (EnKF) is used when the standard Kalman Filter is computationally too expensive in large-scale applications \citep[see][for a comparison between the EnKF and 4D-VAR]{Kalnay_TA_2007}.
If the problem is not large (such as the two-dimensional retrieval over a domain of $50\times50$~km$^2$), the assimilation step can be easily solved with 3D-VAR and the Kalman Filter for the forecast step in which the covariance of the prior is not constant but dependent on time. 
In the following paragraphs we describe the building blocks of our variational Kalman filter framework.

\subsection{Problem description}
The equations that describe our problem are the following:
\begin{subequations}
  \label{eq:model_static}
  \begin{alignat}{2}
&\mathbf{x}_{t}= m_{t-1}(\mathbf{x}_{t-1})+\boldsymbol{\eta}_{t},\\
&\mathbf{y}_{t}=h(\mathbf{x}_{t})+\boldsymbol{\epsilon}_{t},
\end{alignat}
\label{prob}
\end{subequations}
where 
$\mathbf{x}_t$ is the state vector of dimension $N_x$, i.e., the vector of variables we want to estimate, $m_t$ is the non-linear \textit{forecast model} used for the forecast step (giving a forecasted state) from time step $t$ to $t+1$, and $\boldsymbol{\eta}_t$ is the model error. We assume that the model error is additive and Gaussian with zero mean and covariance $\mathbf{Q}$: $\boldsymbol{\eta}_t\sim\mathcal{N}(0,\mathbf{Q})$.  $\mathbf{y}_t$ is the observation vector of dimension $N_y$, i.e., the vector of measurements at time $t$, $h$ is the non-linear \textit{forward model} that project the state vector $\mathbf{x}_t$ into the observation space at time $t$, and  $\boldsymbol{\epsilon}_t\sim\mathcal{N}(0,\mathbf{R})$ is the associated error, where $\mathbf{R}$ is the covariance matrix of the observation errors. 
The best possible estimate of the state at $t$ is obtained using the forecasted state as prior and the newly available observations (assimilation step) giving an updated state (also called analysis state). In the following, the forecast and analysis will be denoted using the superscripts ``f'' and ``a''.

\subsection{Assimilation step}
\label{sec_ass_step}
The assimilation step is based on the variational approach proposed by \cite{Bianchi_JHM_2013b}.
The state vector we want to retrieve is given by
\begin{linenomath}
\begin{equation}
\mathbf{x}^a_t=
\begin{pmatrix}
\ln(\mathbf{r_t})\\
\ln(\boldsymbol{\alpha}_t)
\end{pmatrix},
\label{an_state}
\end{equation}
\end{linenomath}
that is the logarithm of the rain rate values above each pixel and of $\alpha_i$, the prefactor of the k-r power-law~(\ref{att}) above each pixel covered by the links. The exponent of this power law can be reasonably supposed constant and close to 1 \citep{Bianchi_JHM_2013b}.

To obtain the forward rain gauge, radar and microwave link observations, after exponentiating the state values, we used the following forward relations:
\begin{subequations}
\label{eq:model_static2}
\begin{alignat}{2}
&\mathbf{r}_t^g=\mathbf{r}_{t}^f+\boldsymbol{\epsilon}^g_t,\\
& \mathbf{A}_t^l=\sum^{l}_{i=1} u_i \alpha^f_{t_i} \big(r^f_{t_i}\big)^{\beta}+\boldsymbol{\epsilon}^l_t,\\
& \mathbf{r}^r_t= \mathbf{r}_{t}^f+\boldsymbol{\epsilon}^r_t,
\end{alignat}
\end{subequations}
where $u_i$ is the length of the link path in the $i$th pixel, so $\sum^l_{i=1} u_i=L$, with $L$ the total length of the link path; the observation vectors $\mathbf{r}^g_t, \mathbf{A}^l_t$ and $\mathbf{r}^r_t$ are respectively the intensity recorded by the rain gauges, the microwave link path integrated attenuations and the radar-derived rain rates. 
The vectors $\boldsymbol{\epsilon}^g_t, \boldsymbol{\epsilon}^l_t$ and  $\boldsymbol{\epsilon}^r_t$ represent the observation errors \citep[similarly to][]{Bianchi_JHM_2013b}.
The background state estimate $\mathbf{x}^f$ is the rain rate estimate at the previous time step and forecasted (see Section \ref{forecast_step}). 

The analysis state $\mathbf{x}^a_t$ of Eq.~(\ref{an_state}) is obtained by solving the following optimization problem \citep{Bianchi_JHM_2013b}:
\begin{equation}
\mathbf{x}_{k+1} = \mathbf{x}_k + \mathbf{K} \big[\mathbf{H}' \mathbf{R}^{-1} (\mathbf{y}_t-\mathbf{H}'(\mathbf{x}_k))- (\mathbf{P}^f_{t})^{-1}(\mathbf{x}_k-\mathbf{x}^f_t)\big],
\label{ass_step}
\end{equation}
where the matrix $\mathbf{K}$ is the inverse of the Hessian and is given by
\begin{equation}
\mathbf{K}=[(\mathbf{P}_t^f)^{-1}+(\mathbf{H}')^{T}\mathbf{R}^{-1}\mathbf{H}']^{-1}.
\end{equation}
The matrix $\mathbf{P}_t^f$ is the covariance matrix of the prior state error and is described in Section~\ref{forecast_step}, $\mathbf{H}'$ is the Jacobian due to the linearisation of the forward model $h$ in Eq.~(\ref{prob}).

At the first time step $t_0$, there is no previously forecasted rain rate values to use as prior, so we consider as a first guess of the rain rate the radar-derived rain rate. In this way, the convergence is faster than with climatological values and we do not have to put a too low penalty on departures from the background with the risk to give too much weight to the extreme values of the observations. 
The error covariance matrix of the first guess $\mathbf{B} = \langle(\mathbf{x} - \mathbf{x}^b)^T (\mathbf{x} - \mathbf{x}^b)\rangle$ is:
\begin{equation}
	\mathbf{B} =
	\begin{pmatrix} 
		B_{11}  & B_{12} & \cdots     & B_{1s}    & 0               & \cdots &  0  \\
		B_{21}  & B_{22} & \ddots     & \vdots     & \vdots          & \ddots & \vdots \\ 
		\vdots  & \ddots & \ddots     & B_{(s-1)s} & \vdots          & \ddots & \vdots \\
   	   B_{s1}  & \cdots & B_{s(s-1)} & B_{ss}     & 0               & \cdots & 0 \\
   	   0       & \cdots & \cdots     & 0          & B_{(s+1)(s+1)}  & \cdots & 0\\
		\vdots  & \ddots & \ddots     & \vdots     & \vdots          & \ddots & \vdots  \\
		0       & \cdots & \cdots     & 0          & 0& \cdots & B_{(s+p)(s+p)}\\ 
	\end{pmatrix},
	 \label{matrix_B}
\end{equation}
where $s$ is the total number of pixels over which the rain rate is retrieved (for example $s=400$ in our application)
and $p$ is the total number of pixels covered by the microwave links (which varies from one application to another), so the number of retrieved values is $N_x= s+p$. The path of each link has been divided into segments defined by the length of its path over each pixel. The path-averaged rain rate is computed as the sum of the different pixel values weighted by the respective length of each segment.
The first quadrant of the covariance matrix of the state is given by
\begin{linenomath}
\begin{equation}
B_{ij} =B_{ii} \exp \Big(-\frac{ d_{ij}}{d_0}\Big),
\label{cor_dist}
\end{equation}
\end{linenomath}
where $B_{ii}$ is the radar error and $d_{ij}$ is the distance between the pixel $i$ and $j$, with $i,j=1,...,s$, and $d_0$ is the fitted e-folding distance. 
The diagonal of the second quadrant represents the $\ln(\boldsymbol{\alpha})$ errors. 
The covariance matrix of the observation errors $\mathbf{R}$, is given by 9 blocks (except at the first step since the radar is not used as an observation but as a first guess, because of data availability, so the matrix in this case would be composed just by 4 blocks):
\begin{equation}
\mathbf{R} =
 \begin{pmatrix}
\sigma^2_{r_{1}}    & \hdots            & 0               & 0   &\hdots     &0   &0  &\hdots  & 0 \\
\vdots               & \ddots       & \vdots    &     \vdots    &   \ddots     & \vdots   & \vdots &\ddots & \vdots   \\
0              & \hdots         & \sigma^2_{r_{s}}  & 0        &    \hdots    &     0   &  0 &  \hdots & 0 \\

0                   & \hdots  & 0           & \sigma^2_{g_1}   & \hdots  &     0              &  0     & \hdots & 0\\
\vdots            &\ddots   & \vdots   & \vdots       & \ddots                 & \vdots   & \vdots &\ddots & \vdots   \\
0                      & \hdots  & 0           & 0     & \hdots    & \sigma^2_{g_{d}}  &  0   &     \hdots   & 0 \\

0                  &\hdots           & 0     & 0               & \hdots      & 0           & \sigma^2_{l_1}    &\hdots      & 0    \\
\vdots          &\ddots            & \vdots    & \vdots        & \ddots     &\vdots     & \vdots              &\ddots & \vdots   \\
0                 &\hdots            & 0    & 0               & \hdots     & 0    & 0 & \hdots  & \sigma^2_{l_{q}}  
 \end{pmatrix},
 \label{matrix_R}
\end{equation}
where $d$ is the number of rain gauges and $q$ the number of microwave links. See Section~\ref{data}.\ref{sec_parametrization} for more details on the assumptions made.

The inverse problem (\ref{ass_step}) has been solved by applying the Gauss-Newton method, which is stable for systems that are only weakly non-linear. The Gauss-Newton method (and the Levenberg-Marquardt method which can be alternatively implemented when the Hessian is not invertible since it checks to see that each step does reduce the cost function and if not it produces a shorter step size that is also moved in the direction of steepest descend) performs the minimization using also the curvature information which enables the latter to converge with far fewer iterations. The down side is that the computational cost and memory usage for each iteration increase more rapidly then the size of the state vector, but if the problem is small the Gauss Newton is well suited. Moreover the Gauss-Newton method provides the solution error by simply inverting the Hessian.
The incremental procedure used to make the 3D-Var more efficient, in which a non-linear assimilation problem is replaced by a sequence of approximate linear least-square problem \citep{Courtier_1994}, has been shown to be equivalent to an approximate Gauss-Newton method and its convergence has been established \citep{Lawless_2005, Gratton_2007}.

\subsection{Forecast step}
\label{forecast_step}
In this section we describe in detail the \textit{forecast step}:
\begin{equation}
\mathbf{x}_{t}^f= m_{t-1}(\mathbf{x}_{t-1}^a)+\boldsymbol{\eta}_{t-1}.
\label{for_eq}
\end{equation}
The forecast model of the retrieved vector $\boldsymbol{\alpha}$ is assumed to be described by the identity matrix, so for this vector there is no propagation error. The errors in the $\boldsymbol{\alpha}$ estimates are described by the inverse of the Hessian. In reality the variability of the prefactor $\boldsymbol{\alpha}$ depends on the features of the link (frequency and polarization) but also on the drop size distribution and temperature. 
Nevertheless we do not attempt to propagate the  $\boldsymbol{\alpha}$ errors in this study because of the complexity to distinguish the part of inaccuracy that is microwave link dependent, and the part of inaccuracy  due to rain. Thus we will here focus exclusively on the propagation of the rain rate field and its error.

Under the Lagrangian assumption, that is the total intensity does not change during the forecast, the function $m$ of Eq.~(\ref{for_eq}) is an advection model, described by an hyperbolic partial differential equation. Denoting $\mathbf{u}=(u,v)$ the velocity vector, and assuming it remains constant in space and time during the lead-time (i.e., $\nabla \mathbf{u}=0$), the advection equation for a conserved quantity like the scalar field \textbf{r} (rain rate) is described by the following equation:
\begin{equation}
\frac{\partial r}{\partial t} + u \frac{\partial r}{\partial x} +v\frac{\partial r}{\partial y}= 0.
\label{adv_eq}
\end{equation}
There are different ways to discretize Eq.~(\ref{adv_eq}), we use here Total Variance Diminishing (TVD), which is a very accurate scheme \citep{Thuburn_MWR_1996, Akella_2006}. TVD scheme produces no spurious wiggles and is conservative (except at the boundaries). The TVD property arises by doing a weighted average between the Lax-Wendroff Flux and the Upwind Flux between cells.

The velocity components $(u,v)$ of Eq.~(\ref{adv_eq}) are computed using a tracking code in which a Fast Fourier Transform (FFT) is used to compute the cross-correlation and the maximum of this is the best displacement \citep{Panofsky_1968}. 
The short-term rain rate forecasts are produced by assuming Lagrangian persistence, similarly to \cite{Germann_MWR_2002} and \cite{Mandapaka_WF_2011}, that is the radar data are used to extrapolate the velocity and advect the rain rate fields, and the total intensity does not change \citep{Zawadzki_JAM_1973}.

To compute the propagated error covariance matrix $\mathbf{P}_t^f$ given in  Eq.~(\ref{prop_mat}), we used the linear Leapfrog scheme to discretize $m_{t}$, which is second order in space and time \citep{Vukicevic_2001}.

\subsection{Variational Kalman filter scheme}

To summarize, the variational Kalman filter steps implemented in this work are the following:
 
\begin{itemize}

\item[1.] {Forecast step:}
\begin{equation}
\mathbf{x}^f_{t}=\mathbf{M}_{t-1} \mathbf{x}^a_{t-1},
\end{equation}
\begin{equation}
\mathbf{P}^f_{t}=\mathbf{M}_{t-1} \mathbf{P}^a_{t-1} \mathbf{M}_{t-1}^T+\mathbf{Q}.
\label{prop_mat}
\end{equation}
\item[2.] {Assimilation step:}
\begin{equation}
\mathbf{x}_{k+1} = \mathbf{x}_k + \mathbf{P}_t^a \big[\mathbf{H}' \mathbf{R}^{-1} (\mathbf{y}_t-\mathbf{H}'(\mathbf{x}_k))- (\mathbf{P}^f_{t})^{-1}(\mathbf{x}_k-\mathbf{x}^f_t)\big],
\end{equation}
where
\begin{equation}
\mathbf{P}_t^a=[(\mathbf{P}_t^f)^{-1}+(\mathbf{H}')^{T}\mathbf{R}^{-1}\mathbf{H}']^{-1}.
\end{equation}
\end{itemize} 

The analysis state vector $\mathbf{x}^a_t$ is obtained following \cite{Bianchi_JHM_2013b}, the analysis covariance matrix is given by the gain matrix $\mathbf{P}_t^a$. The propagation error $\mathbf{P}_t^f$ of Eq.~(\ref{prop_mat}) is propagated using the linearised forecast model $\mathbf{M}_{t-1}$; $\mathbf{Q}$ is the covariance matrix of the advection model error computed by comparing the advected and recorded radar-derived rain rates, $\mathbf{H}'$ is the Jacobian due to the linearisation of the forward model $h$ in Eq.~(\ref{prob}). This error includes the error due to the advection model and the error due to the Langrangian persistence assumption.
A more detailed description of this filtering can be found in \cite{Lewis_2006} and the notations are conform to \cite{Ide_JMSJ_1997}.

\section{Data and study area}
\label{data}

\subsection{Instruments}

\begin{figure}[t]
	\includegraphics[width=0.5\textwidth]{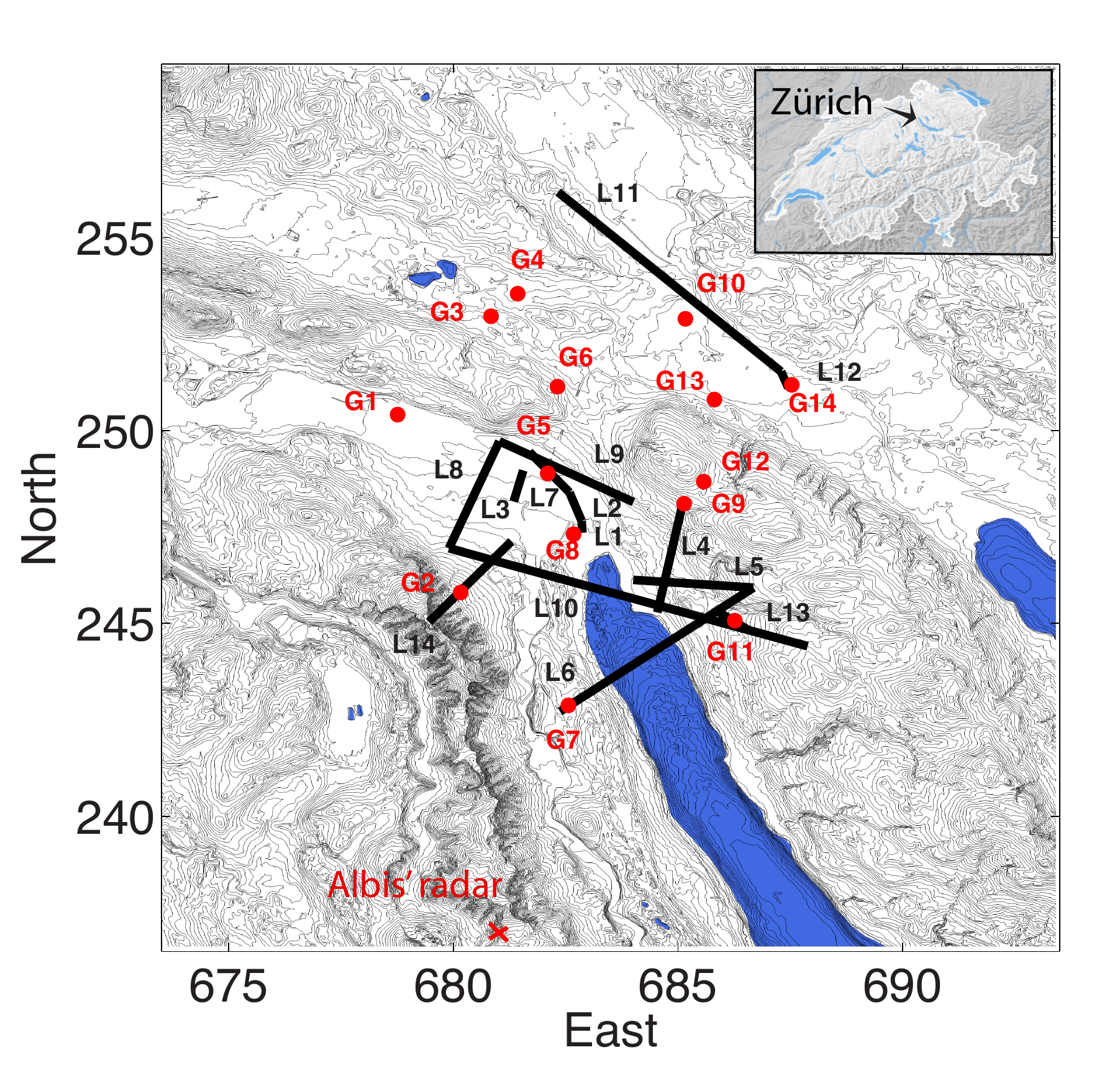}
	\caption{\label{mappe} Locations of the microwave links, rain gauges and Albis' radar (in Swiss grid) in the area of Z\"{u}rich, Switzerland, from \cite{Bianchi_JHM_2013b}.}
\end{figure}

The proposed approach was applied on a data set collected in Z\"{u}rich (Switzerland) in 2009, that has been used in \cite{Bianchi_JHM_2013b}. 
The radar rain-rate fields has a spatial resolution of $1\times 1$~km$^2$ and a temporal resolution of 5 min. The rain rate estimated by the radar network managed by MeteoSwiss was used to obtain the advection field and combined with the rain rate recorded from 14 rain gauges and estimated from 14 microwave links managed by Orange CH (see Figure~\ref{mappe}).

The rain gauge network is composed of 13 tipping-bucket gauges and one weighing gauge with different time resolutions. To make the measurements from rain gauges, microwave links, and radars comparable, all rain gauge measurements have been resampled at 5~min. This resampling assumes a uniform distribution within the original time steps. The value for the new time step was obtained by summing the respective proportion in the covered original time steps.
The microwave links work at 23, 38 and 58~GHz at horizontal or vertical polarization and with a power resolution of 0.1 or 1~dB. The microwave links have been resampled similarly to the rain gauges and pre-processed to remove erroneous records as well. The microwave link characteristics are summarized in Table~\ref{table_mw}. To obtain the attenuation from the RSL, we adopt the simple approach proposed by \cite{Leijnse_WRR_2007_c}: the baseline (attenuation when there is no rain), was defined as the mode of the measured RSL over a sufficiently long period (typically a few months).
Following \cite{Overeem_WRR_2011} the wet-antenna attenuation is set to 1.5~dB.

\begin{table*}[t]
	\begin{center}
		\begin{tabular}{ccccccccccccccc}
			\hline
			Link &1&2&3 &4 &5 &6 &7 &8 &9 &10 &11 &12 &13 &14\\
			\hline
			Freq. &58 & 38 & 38 & 38 & 38 & 23 & 38 & 23 & 23 & 23 & 23 & 58 & 38 &38\\
			Pol.  & V & H & H & V & H & H & H & H & V & V & H & V & H & H\\
			Length & 0.3 &0.8 & 0.8 & 2.9 & 2.7 & 5.4 & 1.4 & 3.0 & 3.4 & 8.4 & 6.8 & 0.5 & 0.8 & 2.8\\
			P. Res. &0.1 &1 &0.1 &0.1 &0.1 &0.1 &0.1 &1 &1 &0.1 &0.1 &0.1 &1 &0.1 \\
			\hline
		\end{tabular}
	\end{center}
		\caption{\label{table_mw} Frequency (GHz), polarization (H or V), length (km) and power resolution (dB) of the 14 microwave links.}
\end{table*}

\subsection{Parametrization}
\label{sec_parametrization}

All measurements errors  are supposed to be uncorrelated. A similar assumption for radar observations has been made by \cite{Hogan_JAMC_2007}. This simplistic assumption is a convenient form for the observation matrix and is common in data assimilation applications. The contribution of the spatial correlation of the retrieved rain field is described by the matrix $\mathbf{B}$ (Eq.~(\ref{matrix_B})) and consequently by $\mathbf{P}$ (Eq.~(\ref{prop_mat})). The rain rate recorded by the rain gauge is assumed to have a representativeness error of $0.58$ in log scale on $1 \times 1$~km$^2$. The error in the radar-derived rain is estimated to be $0.68$ in log scale. The total attenuation errors are fixed at $\Delta A_q = 0.8$ and $1.2$ dB. 
These values together with additional details on the model parametrization of the observation and background errors are described in \cite{Bianchi_JHM_2013b}.
The diagonal elements of the first quadrant of the matrix $\mathbf{B}$ are equal to $B_{ii} = 0.68$ (in log) and they represent the radar error in estimating the rain rate (in log) and $d_{ij}$ is about 1.5 km. 
The diagonal elements of the second quadrant of the matrix $\mathbf{B}$ are such that $\Delta \alpha \approx \Delta \ln(\alpha) \cdot \alpha$ is around 0.13, 0.36 and 0.60 for the link at 23, 38 and 58 GHz respectively.
The first guess values for $\alpha$ and the parameter $\beta$ of the k-R power-low of Eq.~(\ref{att}) are listed in Table \ref{alfa_beta}. The $\boldsymbol{\alpha}$ values have been estimated from DSD measurements and scattering simulation, like in \cite{Bianchi_JHM_2013b}.
\begin{table}[htb!!!]
\begin{center}
\begin{tabular}{ccccc}
\hline
 Freq.   &$\alpha_H$ &$\beta_H$&  $\alpha_V$ &$\beta_V$  \\
\hline
23 GHz  &0.12 &1.06 &0.10 &1.01  \\
38 GHz  &0.29 &0.93 &0.25 &0.91 \\
58 GHz &0.45 &0.81 &0.40 &0.80  \\
\hline
  \end{tabular}
 \caption{\label{alfa_beta} $\alpha$ and $\beta$ values from DSD data and scattering simulation.}  
\end{center}
\end{table}

\section{Results and discussion}

In this section we evaluate the quality of the advection model and of the forecasted rainfall fields.
The improvement of the retrieved rain rate due to the assimilation of rain observations has been demonstrated and quantified in \cite{Bianchi_JHM_2013b} using a cross validation approach where one rain gauge was kept out of the assimilation and used to quantify the error in the prior and in the retrieved rain rate.

\begin{figure}[!hpt]
	\includegraphics[width=0.5\textwidth]{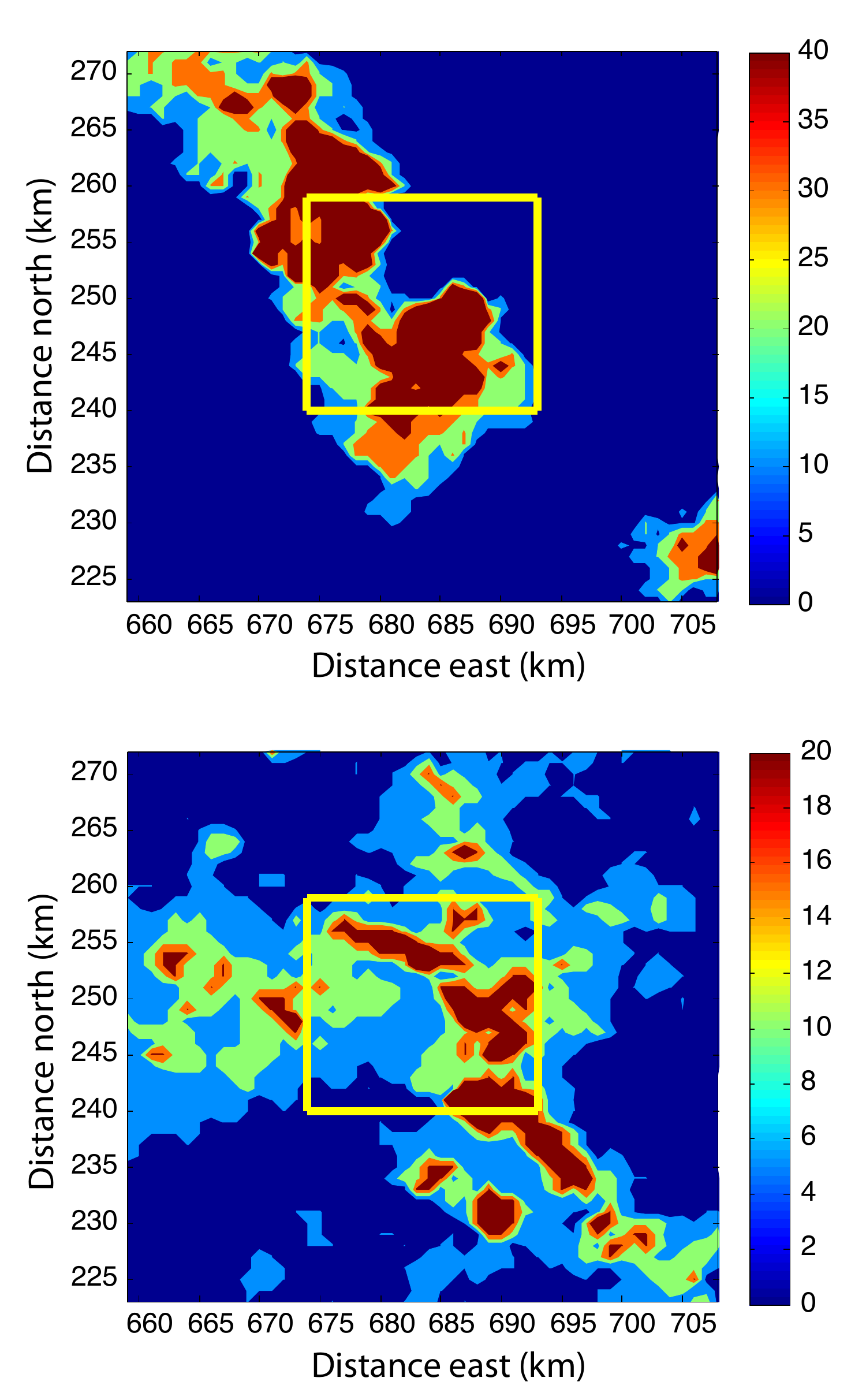}
\caption{\label{radar_obs} Rain rate (mmh$^{-1}$) product derived from operational radar measurements provided by MeteoSwiss, at 14.00~(UTC) on 6 Jun. 2009 (top) and at 8.50~(UTC) on 10 Oct. 2009 (bottom). The yellow square figures the 20$\times$20~km$^2$ area of interest. Note that for readability, the 2 scales are different.}
\end{figure}

The urban area where we tested the algorithm is about 20$\times$20~km$^2$, and is therefore not big enough to reliably compute the velocity field at each time step since we might not have enough rain cells within this relatively small domain.
Thus, the velocity field has been computed using the radar-derived rain over a 50$\times$50~km$^2$ domain (see Figure~\ref{radar_obs}). We do not propagate the covariance matrix of the error when the mean of the retrieval is lower than $0.8$~mmh$^{-1}$ otherwise there are numerical problems to compute the inverse matrices. In such cases we use the inverse of the Hessian of the first time step (that can be seen as a climatological error). 
In addition we also use the same velocity as the earlier time step. 
To test the assumption of uniform velocity over the considered  20$\times$20~km$^2$ area, the divergence of the velocity field has been computed, The divergence values close to zero support the assumption of uniform velocity.

\begin{figure*}[!hpt]
	\includegraphics[width=1\textwidth]{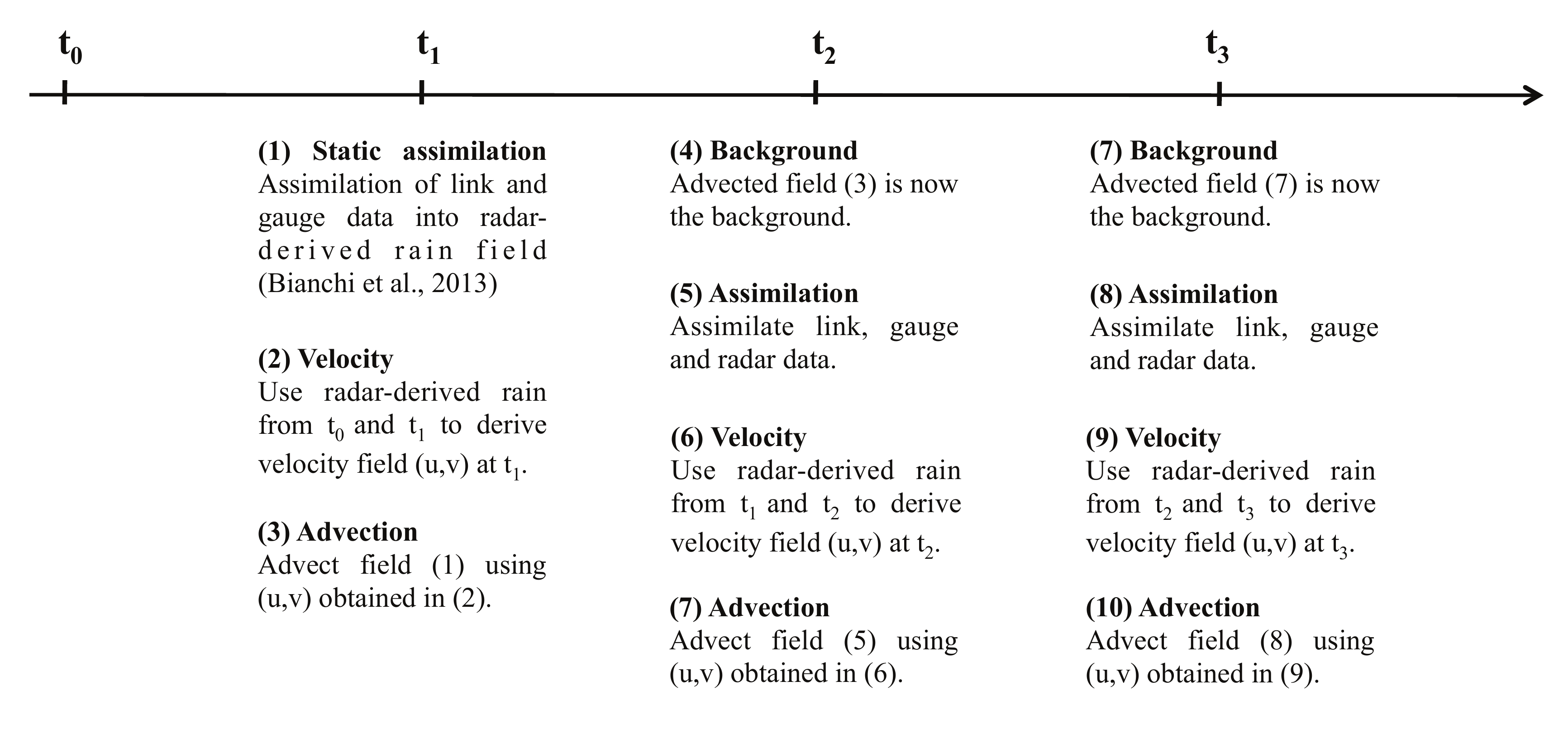}
\caption{\label{nowcast_scheme} Different steps of the proposed nowcasting method.}
\end{figure*}

The results showed in the next two sections correspond to two intense events, one convective event  that occurred on 6 June 2009 and one stratiform event that occurred on 10 October 2009.
Figure \ref{radar_obs} shows the radar-derived rain rates for a representative time step of each event.
The advection of the rain rate field is performed after the assimilation step and the retrieved rain rate field is substituted in the radar-derived rain field of 50 by 50 km$^2$. The flow chart of the nowcast algorithm is provided in Figure \ref{nowcast_scheme}.
  
We computed the Lagrangian error by computing the difference between the intensities where both nowcasted and recorded values indicate rain. To compute the total error, described by the matrix $\mathbf{Q}$ which includes the Lagrangian and advection errors, we computed the difference between the values from the radar product and from the nowcasted fields and removed the values when both were zero. The obtained values are listed in Table~\ref{table_error} and show that the error of the model at 5 min is similar for both the convective and stratiform event, at 30 min the error in the forecasted fields is slightly higher for the convective event (note that the values are provided in log scale).

\subsection{Evaluation of the forecast model}

To assess the performance of the advection model we computed the efficiency, correlation between the radar-derived rain rate advected and the recorded ones at 5, 15, 20, 25 and 30 min, as well as the RMSE for the corresponding indicator fields (i.e., 0 when dry, 1 when rainy).
The efficiency $e$ and correlation $\rho$ are obtained using the following definitions:
\begin{linenomath}
\begin{equation}
e = 1-\frac{\sum^n_{t=1} (r^f_t- r^r_t)^2}{\sum^n_{t=1} (\overline{ r^r_t} - r^r_t)^2},
\end{equation}
\end{linenomath}
and
\begin{linenomath}
\begin{equation}
\rho = \frac{\sum^n_{t=1} (r^f_t- \overline{ r^f_t}) (r^r_t-\overline{ r^r_t})}
{\sqrt{\sum^n_{t=1} (r^f_t-\overline{ r^f_t})^2\sum^n_{t=1} (r^r_t-\overline{ r^r_t} )^2}},
\end{equation}
\end{linenomath}
where $r^f_t$ represents the nowcast rain rate and $r^r_t$ the rain rate estimated by the radar; $n$ is the total number of time steps multiplied by the number of pixels.

To test if the advection model correctly identified the occurrence, this indicator field corresponding to  the nowcasted and recorded values to 0 or 1 when they are respectively dry or wet and then we computed the root mean square error:
\begin{linenomath}
\begin{equation}
RMSE = \sqrt{\frac{\sum^n_{t=1} (o^f_t- o^r_t)^2}{n}}
\end{equation}
\end{linenomath}
where $o^f_t$ represents the nowcast occurences and $o^r_t$ the occurrence measured by the radar.

\begin{figure}[!hpt]
\begin{center}
	\includegraphics[width=0.5\textwidth]{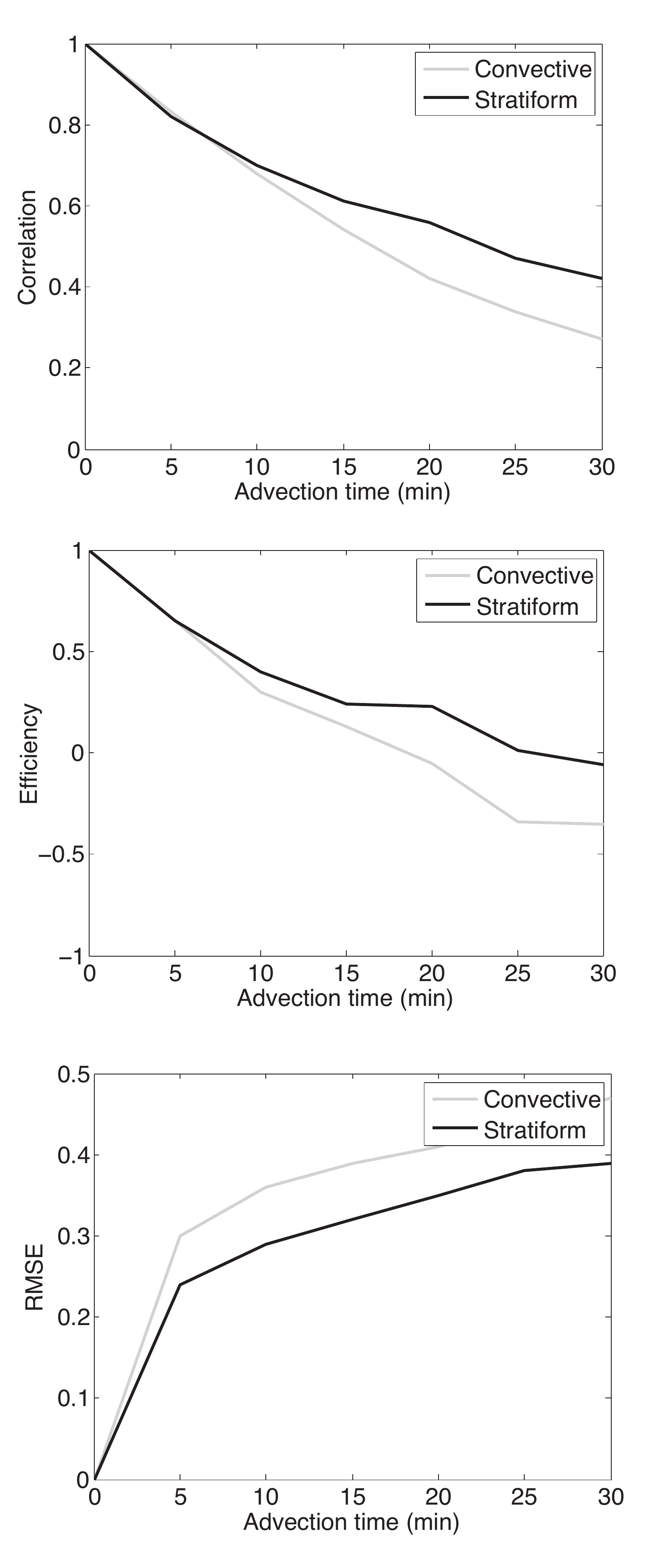}
\caption{\label{fig_statistics2}  Correlation (top), Efficiency (middle) and indicator RMSE (bottom) of the indicator fields as a funciton of the lead-time for the convective and stratiform events.}
\end{center}
\end{figure}

The efficiency, correlation and RMSE values obtained for the convective and stratiform events, for the nowcasted rain rate values are showed in Figure~\ref{fig_statistics2}.
The (simple) advection model exhibits reasonable skills up to 10-15 min for both convective and stratiform events ( $\rho\geq 0.6$; $e\geq 0.3$; $RSME\geq 0.35$) that linearly decrease with the (increasing) lead-time. Beyond 25 min, the performance for the convective case drop faster than for the stratiform one in all criteria. There remains little if any predictive skill for a lead-time of 30 min.
This can be related to the limited extension of the study area, and improving the performance would require data at a larger scale.

Extending our nowcast algorithm to a longer lead-time would require to obtain data at a larger scale, which might be difficult to accomplish given the often limited and patchy distribution of the rain sensors. In addition the assumption of uniform velocity for a lead-time longer of 30 min can be expected to be violated. These factors, combined with an increased computational cost, critical for real time applications, would make it difficult to test the performance of such nowcasts at a longer time scale. 

Focusing on the occurrence prediction, the RMSE values of the convective event are consistently larger than the stratiform ones. At this stage, we have to specify that when we computed the correlation and efficiency we did not consider the pixels when both were dry, in order to avoid biasing the estimation as otherwise the correlation and efficiency would be too high. Similarly when we computed the root mean square error of the occurrences, we do not considered the rain rate cells below 0.8~mmh$^{-1}$ too likely to appear and disappear between successive time steps.

Generally the forecast model shows a good performance for 15~min lead-time. The performance of the algorithm in the convective case decreases rapidly after this lead-time as shown by correlation and efficiency in Figure~\ref{fig_statistics2}, although there is still a good response of the forecast model for the stratiform event until 30~min. The Lagrangian persistence plays an important role in both cases as the RMSE rapidly increases during the first 5~min, more rapidly for the convective case, where after that they both show a similar behaviour.

\subsection{Evaluation of the forecasted rain rate fields}

To assess the performance of the forecasted rain rate fields, the forecast from 5, 10, 15, 20, 25 and 30 min earlier has been compared with the retrieved rain rate at the considered time step. Figure~\ref{fig_forecast_performance} shows the correlation and efficiency values of the nowcasted vs the retrieved rain rates for the two events at different lead times.

\begin{table*}[htb!!!]
\begin{center}
\begin{tabular}{ccccc}
\hline
  &E1 (5 min) & E2 (5 min)  & E1 (30 min) & E2 (30 min)\\
\hline
Lag &0.33 &0.31  & 0.61 &0.53 \\
Tot  &0.34 &0.33   &0.69 &0.58\\
\hline
  \end{tabular}
 \caption{\label{table_error} Lagrangian (Lag) and total (Tot) error (in log) of the nowcast vs recorded at 5 and 30 min, for two events occurred on 6 Jun. (E1) and 10 Oct. (E2) in 2009.}  
\end{center}
\end{table*}

\begin{figure*}[!hpt]
\begin{center}
	\includegraphics[width=1\textwidth]{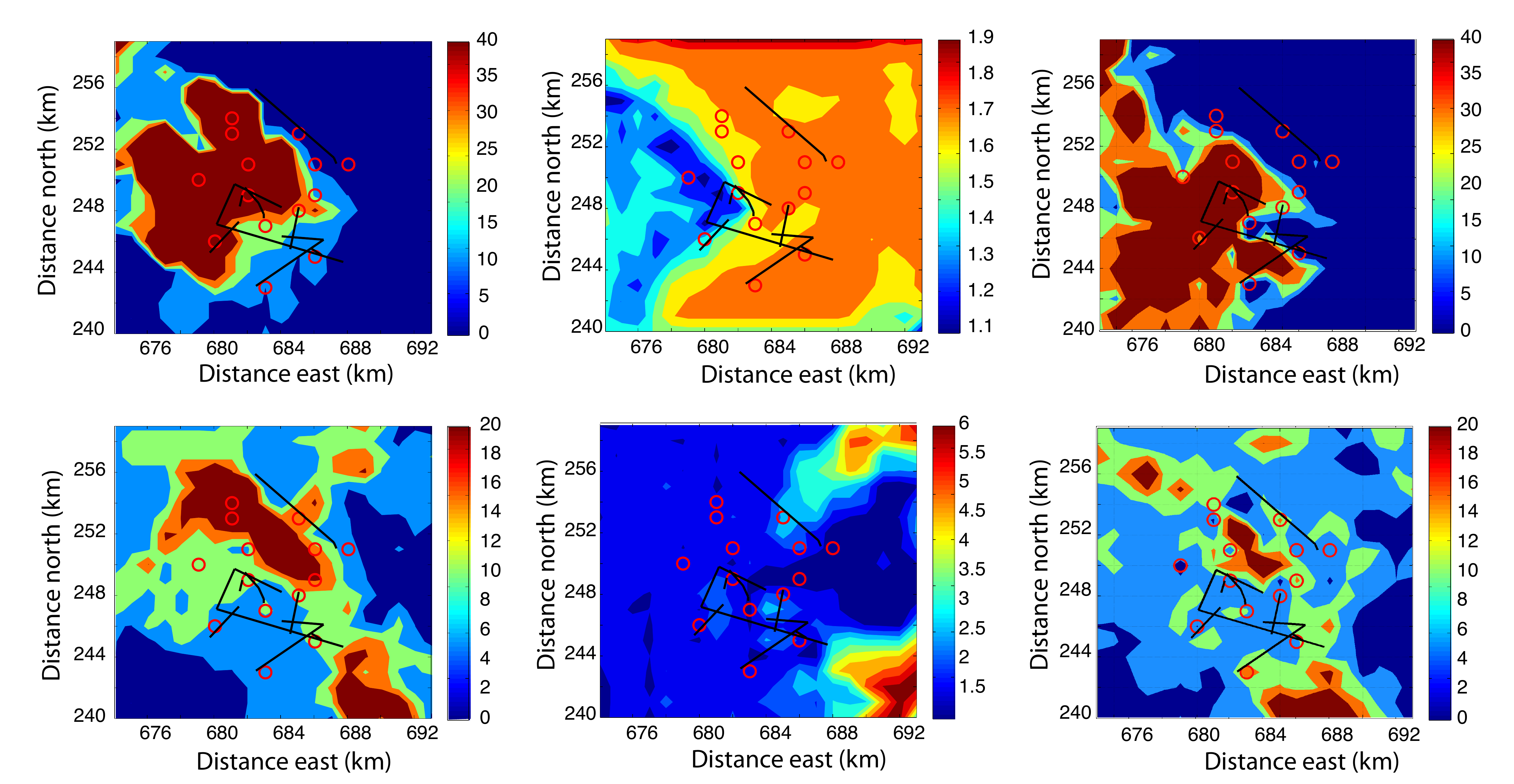}
\caption{\label{nowcast_2D} Nowcast (mmh$^{-1}$), nowcast uncertainty (in log) and retrieval rain rate (mmh$^{-1}$), at 14.00~(UTC) on 6 Jun. 2009 (top) and at 8.50~(UTC) on 10 Oct. 2009 (bottom).}
\end{center}
\end{figure*}

\begin{figure}[!hpt]
\begin{center}
	\includegraphics[width=0.5\textwidth]{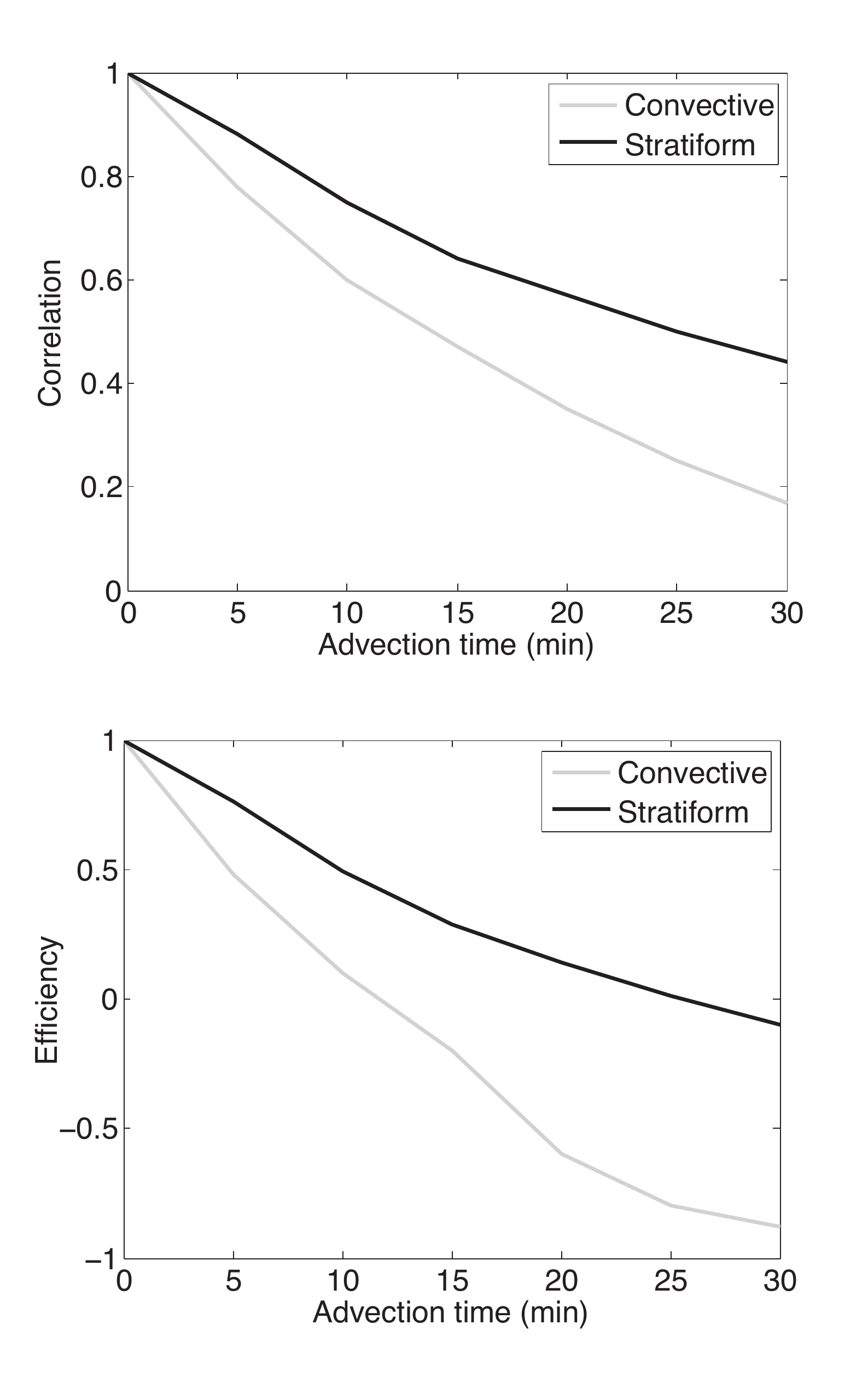}
\caption{Correlation and Efficiency of the nowcast vs retreived rain rate field for the convective and stratiform events.}
\label{fig_forecast_performance}
\end{center}
\end{figure}

Figure \ref{nowcast_2D} shows an example of the rain rate retrieved and forecasted 15 min earlier, with the associated uncertainty, for two rain events.  This lead-time is a good trade-off between having reasonable performance of the advection model and have enough time for warning, particularly in the convective case. The efficiency is about $0.30$ for the stratiform event and $0$ for the convective event. The correlation is about $0.64$ for the stratiform event and $0.47$ for the convective event. This indicates that our assumption are not valid anymore for nowcasting of  convective event further than 15 min in advance. 

\begin{figure*}[!hpt]
\begin{center}
	\includegraphics[width=1\textwidth]{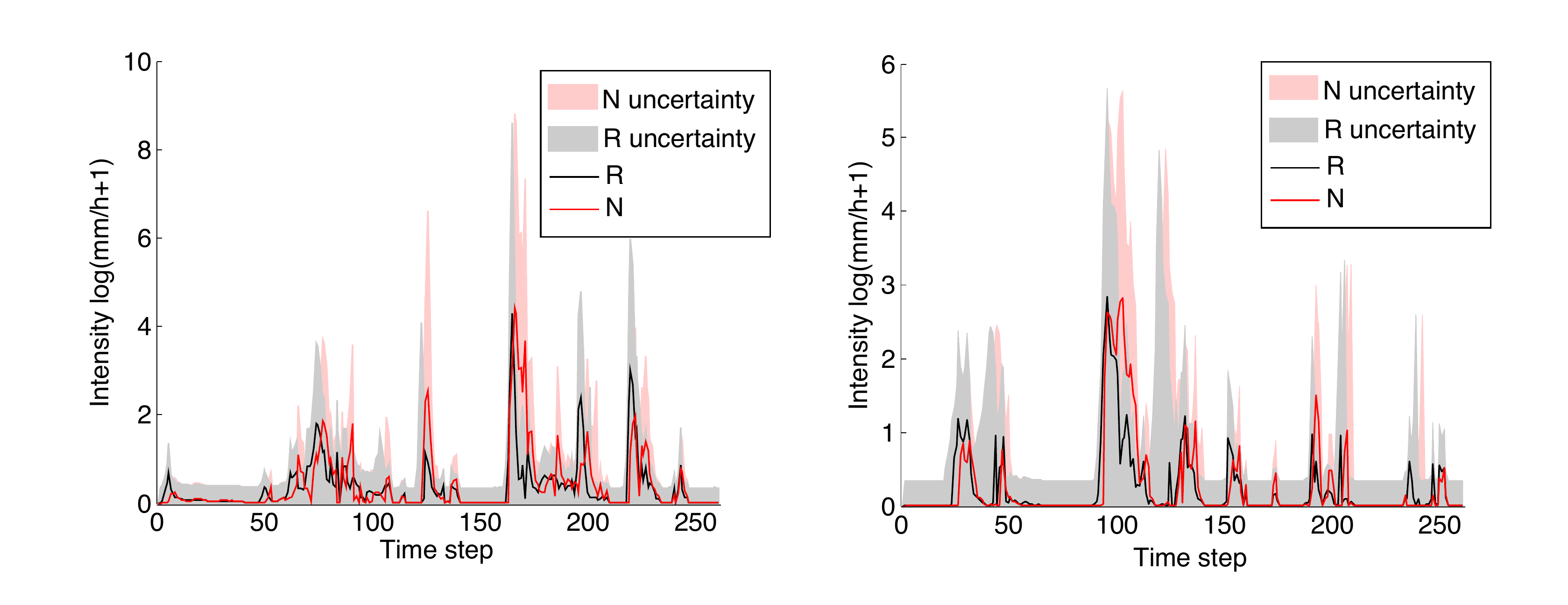}
\caption{\label{nowcast_1D} Nowcasted (N) and retrieved (R) rain rate values with the associated uncertainties (in log scale) on the pixel corresponding to Rain Gauge 3 on 06 Jun. (right) and Rain Gauge 11 on 10 Oct. (left) 2009. Note that the log is estimated on rain rate~+~1~(mmh$^{-1}$), to avoid too low values for very small rain rate values.}
\end{center}
\end{figure*}

Figure \ref{nowcast_1D} shows the rain rate retrieval and forecasted 15 min earlier, with the associated uncertainty, for the convective and stratiform event, at two different rain gauge locations. Similar results have been obtained for the other rain gauges. The retrieved rain rate is included in the uncertainty of the nowcast most of the time, providing a first qualitative validation of the theoretical error provided by the method.

\begin{figure}[ht!]
\begin{center}
	\includegraphics[width=0.5\textwidth]{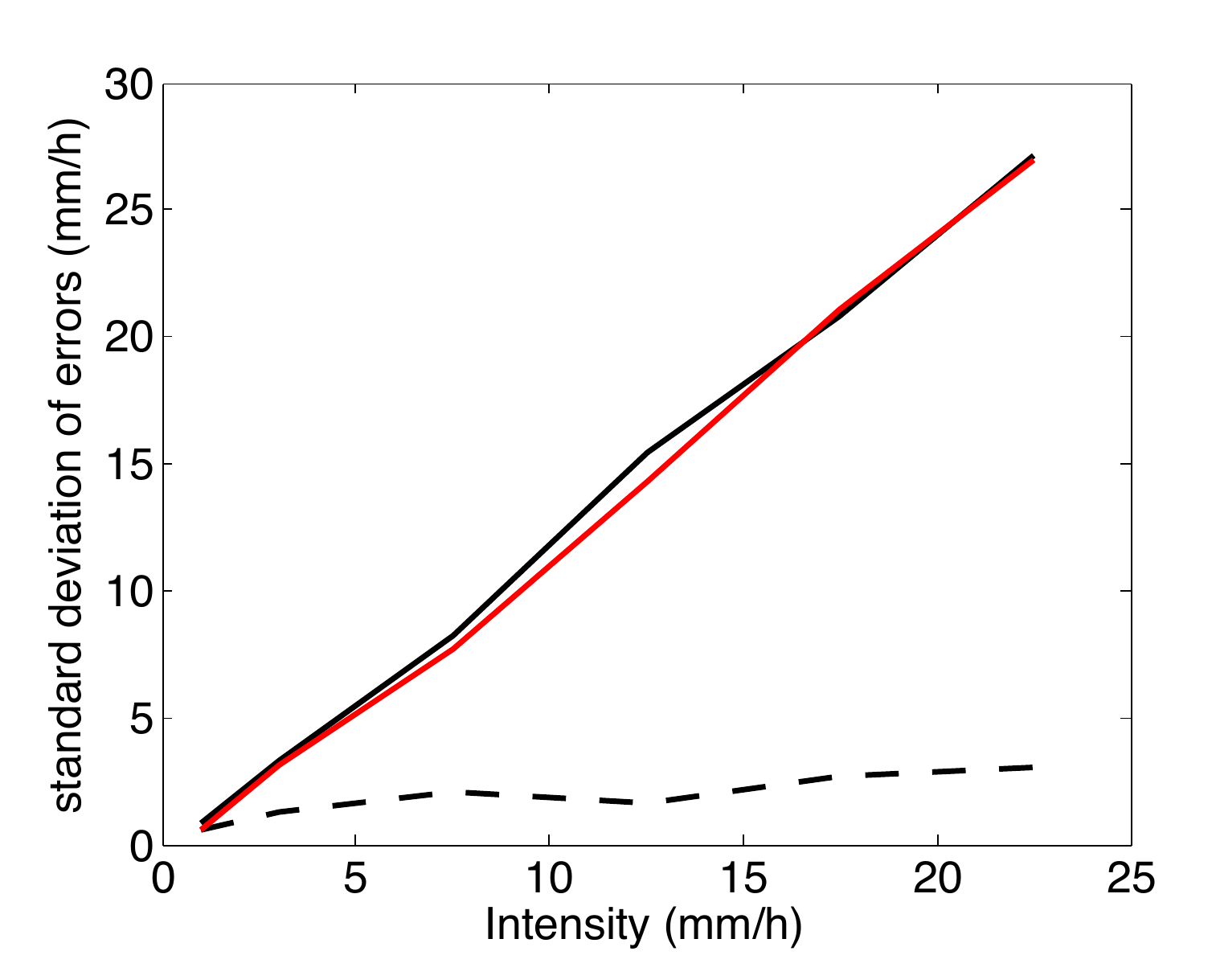}
\caption{\label{fig8} The predicted empirical error (precision) of nowcast (in black) and retrieval (in red) are consistent. The dotted line represents the difference between the retrieval and nowcast (mmh$^{-1}$) and give an information of the accuracy of the rain rate nowcasting. See Table  \ref{Retrieval_error} for the consistency of the retrieval uncertainty with the measurement uncertainty.}
\end{center}
\end{figure}

The significant uncertainty of the nowcast is however consistent with the uncertainty of the retrieval, and dependent on the rain intensity (Figure~\ref{fig8}). The uncertainty of the retrieval is defined as the theoretical error provided by the error covariance matrix, that is the inverse of the Hessian, and can be seen as a measure of the precision of the retrieval. It was found to be around 0.65 (log units)
which is comparable with that of the observations. The error in the radar is 0.68 (log units), which indicates that the uncertainty slightly decrease due to the contribution of the near-by sensors. 
A cross-validation to evaluate the theoretical uncertainty of the retrieval was carried out by keeping out several rain gauges (close to another rain gauge or microwave link) and taken as reference, see Table~\ref{Retrieval_error}. The accuracy of the retrieval, computed as the standard deviation of the difference between the rain gauges and the retrievals were around 0.45 log-units. The difference between the rain gauges and the retrievals is found to be consistent with the spatial representativeness error of the rain gauge, i.e. around 0.48 log units. 
The uncertainties of the retrieval and nowcast are strictly dependent on the reliability and confidence of the measurements. Higher confidence and reliability on the measurements  will naturally lead to a decreased uncertainty in the retrievals.

\begin{table*}[htb!!!]
\begin{center}
\begin{tabular}{ccccc}
\hline
 & 06 Jun &  & 10 Oct. & \\
\hline
 & $\sigma$(Retrieval-Gauge) & $\triangle$Retrieval   & $\sigma$(Retrieval-Gauge) & $\triangle$Retrieval  \\
 \hline
G3 & 0.38 & 0.66 & 0.25 & 0.67 \\
G4 & 0.33 & 0.66 & 0.24 & 0.67 \\
G9 & 0.46 & 0.66 & 0.40 & 0.67 \\
G11 & 0.46 & 0.63 & 0.45 & 0.65 \\
G12 & 0.42 & 0.67 & 0.37 & 0.67 \\
\hline
  \end{tabular}
 \caption{\label{Retrieval_error} Accuracy and precision test of the retrieved rain rate.}  
\end{center}
\end{table*}

\section{Conclusions}

In this study we proposed to use the Variational Kalman Filter approach to obtain nowcasted rain rate fields. This methodology implements a dynamic model of rainfall and is intended for hydrometeorological applications that need nowcasts at high temporal resolution. The Variational Kalman Filter generates limited implementation costs (quadratic convergence and low complexity of the variational scheme), and yields accurate results due to the dynamic error covariance. It is a convenient form for on-line real time processing, it is easy to formulate and to implement. The obtained results correspond to a relatively small urban areas of about 20$\times$20~km$^2$ with a spatial resolution of 1~km and 5 min of temporal resolution. 

Some assumptions are necessary, i.e. the log-normality assumption of the measurements errors and the Lagrangian persistence. 
If there is an evidence that the assumption of the log-normal distribution of errors is likely to be violated, alternative data assimilation techniques should be considered (e.g., particle filter).
The assumption of the Lagrangian persistence appears to be valid up to 20 min, and slightly longer times for stratiform events. In the case of convective events, the performance of the nowcast algorithm decreases rapidly after 15 min, due to rapid development and movement of rain cells. 
The lifetime obtained is however consistent with other nowcast studies. Indeed our method yields results that are comparable to those obtained using other short term forecast approaches at a larger scale. Despite some uncertainty due to the different nowcast methods, radar and data processing, the results presented in this study are consistent with them. 

An important aspect of this work, besides the fact of being able to combine different sensors taking into account their respective uncertainties, is the estimation of the uncertainty associated with the retrieved and nowcasted rain rate fields that are of primary importance for hydrological applications in an urban environment to achieve a better real-time management of sewage systems and limit the impacts of severe overflow on the natural environment. 

Future research could be addressed on more advanced prediction models than simple advection to deal with the rapid development and movement of rain cell especially in the case of convective events to improve local high-resolution forecasting.

\section{Acknowledgments}
We thank Thorwald Stein for providing the tracking code to compute the velocity fields. The radar, microwave link and rain gauge data were kindly provided by J\"org Rieckermann from Eawag. The financial support from the Swiss National Science Foundation for the COMCORDE project is acknowledged (grant CR22D-135551).

\end{document}